\documentclass[reprint,amsmath,amssymb,aps,onecolumn,superscriptaddress]{revtex4-1}

\input{./preambule}

\usepackage{graphicx}% Include figure files
\usepackage{dcolumn}% Align table columns on decimal point
\usepackage{bm}% bold math
\usepackage{xcolor}
\usepackage{amsmath}
\usepackage{comment} 
\usepackage{amssymb}

\begin{document}

\title{Power of Ensemble Diversity and Randomization for Energy Aggregation}  
%Mixing in the Ensemble of Diverse, Thermostatically Controlled Loads} 
%\thanks{A footnote to the article title}

\author{David M\'etivier }
\email{metivier@lanl.gov}
\affiliation{CNLS \& T-4, LANL, Los Alamos, NM}

\author{Ilia Luchnikov}
\email{ilia.luchnikov@skolkovotech.ru}
 \affiliation{Skoltech, Moscow, Russia}

\author{Michael Chertkov}
\email{chertkov@lanl.gov}
\affiliation{CNLS \& T-4, LANL, Los Alamos, NM}
\affiliation{Skoltech, Moscow, Russia}
 \email{chertkov@lanl.gov}

\date{\today}

\begin{abstract}
We study an ensemble of diverse (inhomogeneous) thermostatically controlled loads aggregated to provide the demand response (DR) services in a district-level energy system. Each load in the ensemble is assumed to be equipped with a random number generator switching heating/cooling on or off with a Poisson rate, $r$, when the load leaves the comfort zone. Ensemble diversity is modeled through inhomogeneity/disorder in the deterministic dynamics of loads. Approached from the standpoint of statistical physics,  the ensemble represents a non-equilibrium system driven away from its natural steady state by the DR. The ability of the ensemble to recover by mixing faster to the steady state after its DR's use is advantageous. The trade-off between the level of the aggregator's control, commanding the devices to lower the rate $r$, and the phase-space-oscillatory deterministic dynamics is analyzed. We discover that there exists a critical value, $r_c$, corresponding to both the most efficient mixing and the bifurcation point where the ensemble transitions from the oscillatory relaxation at $r>r_c$ to the pure relaxation at $r<r_c$. Then, we study the effect of the load diversity, investigating four different disorder probability distributions (DPDs) ranging from the case of the Gaussian DPD to the case of the uniform with finite support DPD. Demonstrating resemblance to the similar question of the effectiveness of Landau damping in plasma physics, we show that stronger regularity of the DPD around its maximum results in faster mixing. Our theoretical analysis is supported by extensive numerical validation, which also allows us to access the effect of the ensemble's finite size. 
\end{abstract}

\maketitle

\section{Introduction}

Demand response (DR) is a popular modern way to balance power systems \cite{DR}. It can also be used more broadly to improve control of large engineered systems,  such as natural gas systems,  district heating systems, water systems, traffic systems, and so forth. Many of the infrastructure systems were originally devised assuming a clear separation of roles between,  for example, loads and generators in a power system---loads consume the electric power as they need it without much coordination with generators, while the generators balance the system as fast and accurately as possible to keep the system balanced and running. However, this traditional paradigm is challenged by many modern additions to power systems,  such as wind and solar renewable generation, which involve much more uncertainty and fluctuations than the system experienced in the past, thus making the system less reliable and stable.  The essence of DR is fixing this problem by breaking the traditional split of roles between generators and loads by involving the loads in the system control and coordination. In spite of its relatively short history, DR has now become widely discussed and implemented, mainly through control of large flexible loads (see, e.g., review \cite{13LKRB} and references therein).

In this manuscript, we focus on a type of DR that is less developed---coordination of many small flexible loads to provide DR services to the grid. However, and as has been mentioned in the early papers on the subject \cite{79MB,79CD,81IS,84CM}, the main difficulty in involving small flexible loads in DR is related to the coordination overhead.  Indeed, the benefit of involving a small individual load, say an air conditioner in your apartment, in the DR is too small to make it economically sound.  A viable solution is to consider the small loads in aggregation, as an ensemble, thus introducing a new entity---the aggregator---whose task is to resolve the DR challenge for many (up to tens of thousands) small loads collectively \cite{11CH}. Obviously, the aggregation is economically viable only for a sufficiently large ensemble when the profit from the DR services exceeds expenditures related to communication overhead between the aggregator and the loads and when the resulting load manipulations are not too disruptive to the load's main tasks. In its extreme version, the idea is to rely only on one-way communication between the aggregator and its many consumers/loads via a broadcast so that the communications are minimal and each load receives the same information in real time.  The information would typically contain a command sent to all the devices to switch on or off.

Another complication associated with the collective functioning of many loads, noticed already in  \cite{79MB,79CD,81IS}, is the so-called cold load pickup,  which occurs at the conclusion of the DR service interruption. In this case, sufficiently long involvement of loads in the DR services leads to load synchronization---many thermostatically controlled loads,  typically subject to the bang-bang control switching the loads on/off when they reach the endpoints of the comfort zone, are moving along their path in the  phase space together, thus resulting in long undesirable oscillations of the ensemble cumulative consumption.

Stochastic effects, associated with uncontrolled and short-correlated temporal fluctuations of loads as well as inhomogeneity of loads within the ensemble, destroy the synchronization eventually and the system mixes into a statistically steady state. However,  natural stochasticity and inhomogeneity are typically weak, thus resulting in unacceptably slow mixing. (The slow mixing translates into a delay constraint on the next use of the ensemble in the DR.)

As argued in \cite{12AK,16BMb}, acceleration of mixing can be achieved by adding a controlled random component to the load dynamics. This ``randomize for better mixing'' idea was brought into the context of the aggregator model in \cite{17CC}, where it was suggested to allow the loads to deviate from the bang-bang control. When the loads leave the comfort zone, their state (on or off) is not changed instantaneously  but instead with a delay generated independently by each load according to a Poisson distribution with rate $r$. The rate $r$ is the only parameter that is broadcast to the loads by the aggregator.

The methodology of \cite{17CC}, utilizing the Fokker--Planck (FP) formalism of statistical physics brought into the DR literature in \cite{84CM,09Cal}, was limited to a homogeneous ensemble and to a dynamical load model that was too complex to allow analytic analysis of the mixing conditions. In this manuscript,  we correct for these limitations, thus extending and improving the approach of \cite{17CC}.

\subsection{Main results of the manuscript} 

We study the effects of the load inhomogeneity (which we also call disorder, following the statistical physics jargon) on operations of the ensemble, specifically in terms of the ensemble's ability to recover fast from a perturbation after its use by the aggregator for the DR. The ensemble is assumed to be controlled by an aggregator in a communication-minimal way by sending the same signal switching off/on rate to all the consumers simultaneously. We are mainly interested in the regime where both the control and the ensemble variability are weak. The two main messages of the paper (put here in a colloquial format and then quantified formally later) are as follows:
\begin{itemize}
    \item [(1)] %In the regime with no or small ensemble variability 
    There exists an optimal switching rate corresponding to the fastest recovery. Any deviation (increase or decrease) of the rate leads to a slower recovery. 
    
    \item [(2)] Increase of the ensemble variability is advantageous for faster recovery/mixing. 
    % even in the regime without or with weak (large switching rate) ensemble control.
\end{itemize}

Because temporal evolution of the ensemble is at the core of this manuscript analysis,  let us define relevant timescales and then restate our main results in a more technical way.  We assume that by default (without aggregator), each customer follows a standard bang-bang operation---switching on (off) the cooling device when the temperature exceeds (becomes less than) a preset threshold. We assume that the outside temperature is significantly higher than the switch-on threshold, thus resulting in cycling of the device with its natural timescale $\tau$. The aggregator changes this natural cycling by requesting the consumers to switch on/off with a random delay distributed according to a Poisson distribution with rate $r$. (Each device is assumed equipped with a random number generator.) By default, i.e., without aggregator control, $r=+\infty$. Weak aggregator control means that $r\tau\gg 1$.  To account for variability within the ensemble, one assumes that devices may have slightly different $\tau$. Formally, one considers $\tau$ as the disorder (variability) parameter distributed according the disorder probability distribution (DPD), $g(\tau)$, characterized in terms of its typical value (mean), $\tau_0$, and the distribution width, $\Delta$. In the following we define and consider four different forms of $g(\tau)$---Gaussian, Lorentzian, Laplace, and finite-support uniform---parameterized by $\tau_0$ and $\Delta$, however always assuming (analyzing the disorder case) that the typical control is weak, i.e., $r\tau_0\gg 1$, and that the disorder is also weak, i.e., $\Delta \ll \tau_0$. 

\begin{figure}
        \includegraphics[scale=0.65]{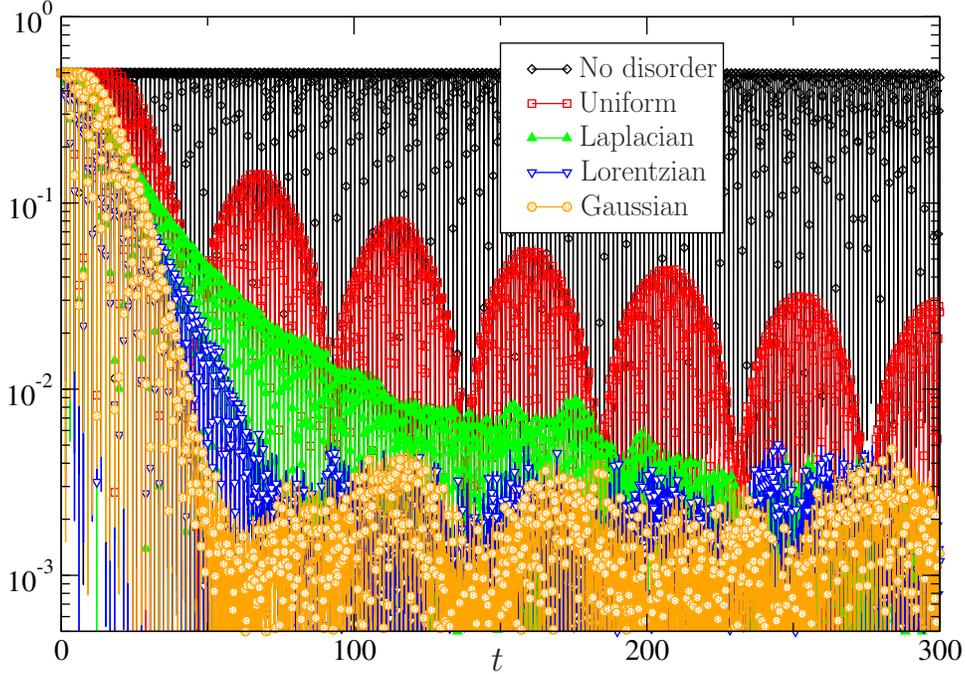}
        \caption{Results of particle simulations providing a representative comparison of the four types of disorder. We chose here the most dramatic case with almost no (additional) control by the aggregator where in the case of no disorder (all consumers are the same) the system does not mix.  Introduction of disorder (diversity in the aggregator's portfolio) results in the decay,  which is most impressive in the case of the Gaussian disorder,  still fast (exponential) in the case of the Lorentzian disorder, and much slower (but still noticeable, $1/t^2$ and $1/t$) in the case of the Laplacian and uniform finite-support disorder. We show here $|{\cal N}_{\uparrow}(t)-1/2|$ vs time, and the model parameters chosen for the comparative illustration are $\tau_0=3$, $r=100$, $\Delta=0.1$, $-\xd=\xu=1$, and $N=10^5$. (See Section \ref{sec:simu} for more details, e.g., for comparison of theory and simulations.)}
        \label{Comparison_diso}
    \end{figure}

With the timescales and two small dimensionless parameters, $(r\tau_0)^{-1}$ and $\Delta/\tau_0$, defined, we are ready to provide the following more technical, still qualitative but intuitive, explanations for our main results.
\begin{itemize}
    \item [(1)] When $r=\infty$, the system does not decay and temporal evolution of the probability distribution function (PDF) of a device temperature, $x$,  averaged over the ensemble shows a periodic behavior in time, $\sim \exp(\pm i\lambda_I t)$, with the period $1/\lambda_I=\tau/(2\pi)$. %=(x_\uparrow-x_\downarrow)/u$. 
    Decrease of $r$ leads to decrease of $\lambda_I$ and simultaneous increase (from zero at $r=\infty$) of the decay rate, $\lambda_R$. In this oscillatory with a decay regime, temporal behavior of the correction to the stationary probability distribution becomes $\sim \exp(-\lambda t)$, $\lambda=\lambda_R\pm i\lambda_I$ $\lambda_I\neq 0$ and $\lambda_R>0$, where $\pm$ reflects emergence of two complex-conjugated solutions.  At a certain critical value, $r=r_c$, $\lambda_I$ becomes zero, i.e., the two complex-conjugated solutions merge into one (degenerate) solution such that close to the merging point $\lambda=\lambda_c(1\pm c \sqrt{1-r/r_c}+O(1-r/r_c))$, where $c=O(1)$ and $\lambda_c$ is the critical value of $\lambda_R$ achieved at $r=r_c$. The main conclusion of this straightforward qualitative estimate is that the lowest of the two eigenvalues (corresponding to $\pm 1\to -1$ and thus to the slowest asymptotic decay) achieves its maximum as a function of $r$ at $r_c$. %(See Appendix ? for additional discussion of the oscillatory analogy.)
    
    \item [(2)] In the default regime (no aggregator control), a set of devices with exactly the same $\tau$,  i.e., when we set $\Delta$ to zero, would not mix at all, i.e., correction to the stationary probability distribution oscillates and does not decay. Introduction of a small but finite $\Delta$ results in a decay that is largely controlled by %the ``regularity" of, 
    $g(\tau)$ in the vicinity of its maximum, i.e., at $\tau\approx \tau_0$. Specifically, decay of the temperature probability to its stationary value in time is controlled by the shifted Fourier transform of the DPD, $\int d\tau g(\tau) \exp(\pm 2\pi i t/\tau)\approx \exp(\pm 2\pi i t/\tau_0)\int d\varsigma g(\tau_0+\varsigma)
    \exp(\mp 2\pi i \varsigma t/\tau_0^2)$. Obviously, details of the decay depend on the shape of the DPD. Of the four model DPDs considered in this manuscript, the Gaussian DPD results in the fastest decay (shifted Gaussian in time), the Lorentzian DPD is a bit slower (exponential in time), and the Laplacian DPD and uniform finite-support DPD are the slowest, with asymptotic $1/t^2$ and $1/t$ decays, respectively. This hierarchy,  illustrated in Figure~\ref{Comparison_diso}, and the Fourier-transform interpretation suggest that the speed of decay is linked to the regularity of the DPD around its central part. (This phenomenon is reminiscent of the mathematically similar analysis of the Landau damping in plasma physics described by the Vlasov equation [see \cite{villani_landau_2010,mouhot2011} and references therein]. Specifically, we refer here  to the fact that the regularity of the initial velocity distribution influences the Landau mixing/damping speed).
\end{itemize}

These main focal points are detailed and extended in the remainder of the manuscript. Models of the statistical ensemble and of the ensemble inhomogeneity are formulated in Section \ref{sec:formulation}. The basic model of the homogeneous ensemble is analyzed in Section \ref{sec:basic}. Effects of the disorder/inhomogeneity are studied in Section \ref{sec:disorder},  where we also compare analytic and numerical results.  Section \ref{sec:conclusion} is reserved for conclusions and discussion of the path forward.

\section{Formulation of the problem}
\label{sec:formulation}

One characterizes a load by the continuous parameter, $x$, standing for the temperature, and by the discrete/binary parameter, $\sigma=\pm 1$, indicating whether the air conditioning system/device of the load (one considers cooling for concreteness) is switched on, $\sigma =+1$, or off, $\sigma=-1$. Conditioned to $\sigma$, the dynamics of $x$ follow the deterministic rule
\begin{eqnarray}
\frac{dx}{dt}=v(x,\sigma),
\label{eq:general_x_model}
\end{eqnarray}
where $v(x|\sigma)$ describes the rate of temperature change as a function of the current temperature, $x$, conditioned to the state of the load's air conditioning device (later in the text referred to simply as ``device''). 
Our basic model is
\begin{eqnarray}
    \mbox{\underline{Basic Model}}:\quad v(x,\sigma)=\left\{\begin{array}{cc} -u, & \sigma=+1\\ u, & \sigma=-1,\end{array}\right.
    \label{eq:basic_x_model}
\end{eqnarray}
where $u$ is a positive constant. The model is a simplification of a bit richer popular model, e.g., used in \cite{17CC}, where $u$ is not a constant as in Eq.~(\ref{eq:basic_x_model}) but a linear function of $x$.  
% We choose here a simpler model guided by considerations of (a) simplicity; and (b) practical relevance in the standard case where the comfort zone of a device is sufficiently far from the temperature fixed points where the customer is in a balanced state with the environment (when the device is switched on and off respectively). %(We will also discuss other, more complex, models later in the text.)
$\sigma$ in Eqs.~(\ref{eq:general_x_model},\ref{eq:basic_x_model}) is modeled as the following Markovian binary (two-level) stochastic process: 
\begin{eqnarray}
    \forall t:\quad \sigma(t+dt)=\left\{\begin{array}{cc} \sigma(t), & x\in [x_\downarrow;x_\uparrow]\mbox{ or otherwise with probability } 1-rdt 
    \\ -\sigma(t), & x\notin [x_\downarrow;x_\uparrow]\mbox{ with probability } rdt \end{array}\right., 
    \label{eq:sigma_model}
\end{eqnarray}
where $dt$ is the time step (of the properly discretized continuous time limit), $r$ is the rate of exponential (Poisson) switching, and $x_\uparrow,x_\downarrow$ marks the size of the temperature band within which no switching occurs, $x_\downarrow<x_\uparrow$.

\begin{figure}
    \includegraphics[scale=0.65]{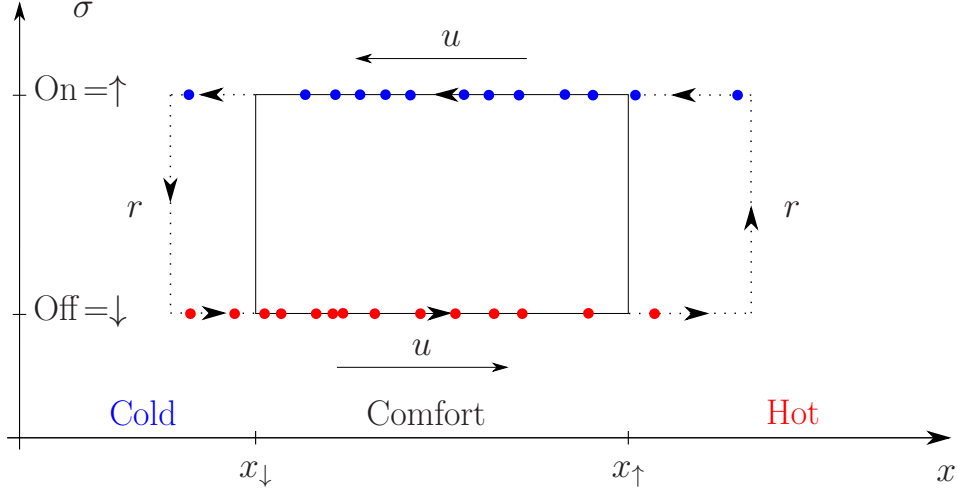}
    \caption{ \label{fig:cycle} Schematic illustration of an instantaneous device distribution and dynamics in the $(x,\sigma)$ phase space, where $x$ is the (continuous) temperature, and the discrete (two-level) designation, $\sigma=\pm $,  marks whether the device (air conditioner) is switched on, $\sigma=+1$, or off, $\sigma=-1$.}
\end{figure}

As set above, the basic model has two timescales: one describing deterministic evolution, $\tau= 2(x_\uparrow-x_\downarrow)/u$, which is the time it takes for a device to make a full cycle through the combined $(x,\sigma)$ phase space illustrated in Figure ~(\ref{fig:cycle}), and $1/r$, the typical time of a stochastic jump from $\sigma=+$ to $\sigma=-$ or vice versa.

Notice  that the selection of the model in Eq.~(\ref{eq:basic_x_model}) as the basis for this manuscript analysis is dictated not only by its realism but also by considerations of simplicity and our ability to derive analytic results. Specifically, for the case of the asymptotic uniform ensemble consisting of the infinite number of devices with the same characteristics (the same $u$) and following the same switching protocol (the same $r$), we are interested in computing and analyzing the evolution in time of the PDFs  governed by the system of coupled FP equations following directly from the model definition, given by Eqs.~(\ref{eq:general_x_model},\ref{eq:basic_x_model},\ref{eq:sigma_model}):
\begin{eqnarray}
    && \left(\partial_t\begin{pmatrix} 1 & 0\\ 0 & 1\end{pmatrix} -\L\right)P(x|t,\tau,r)=0,\quad P(x|t,\tau,r)\doteq\begin{pmatrix} P_{\uparrow}(x|t,\tau,r)\\ P_{\downarrow}(x|t,\tau,r)\end{pmatrix},\label{eq:FP_basic}\\ && \L\doteq
     u\partial_x \begin{pmatrix} 1 & 0\\ 0 & -1\end{pmatrix}
     -r
     \begin{pmatrix} 
         \theta(x_\downarrow -x) & - \theta(x-x_\uparrow)\\ -\theta(x_\downarrow -x) & \theta(x-x_\uparrow)
     \end{pmatrix},
    \label{eq:FP_operator}
\end{eqnarray}
where $\theta(y)$ is unity if $y>0$ and zero otherwise. We are seeking a properly normalized solution of Eq.~(\ref{eq:FP_basic}):
\begin{eqnarray}
{\cal N}_\uparrow(t,\tau,r)+{\cal N}_\downarrow(t,\tau,r)=1,\quad {\cal N}_{\uparrow,\downarrow}(t,\tau,r)\doteq \int dx P_{\uparrow,\downarrow}(x|t,\tau,r),
        \label{eq:densities}
\end{eqnarray}
where ${\cal N}_{\uparrow,\downarrow}(t,\tau,r)$ counts proportions of devices that are switched on and off, respectively. 

As shown in Section \ref{sec:basic}, solution of the system of the FP Eqs.~(\ref{eq:densities}) can be presented explicitly as the spectral expansion in terms of the Lambert-W functions for any initial $t=0$ distributions. This analytic expression will allow us to analyze the temporal evolution of the basic homogeneous ensemble in much more detail than \cite{17CC} for a more complex model, with $v(x|\sigma)$ in Eq.~(\ref{eq:general_x_model}) dependent linearly on $x$.

However,  devices contributing realistic ensembles are not necessarily the same in terms of their cooling/heating strength.  To model the ensemble diversity, i.e., non-uniform ensemble, one introduces disorder in $\tau$. We assume that $\tau$ characterizing a device is drawn independently from one of the following four model DPDs: Gaussian, Lorentzian, Laplace, and uniform (finite support)
\begin{eqnarray}
		g_G(\tau|\tau_0,\Delta)& = & \dfrac{1}{\sqrt{2\pi}\Delta}e^{-\frac{(\tau-\tau_0)^2}{2\Delta^2}},\label{eq:g-Gaussian}\\
		g_{Lr}(\tau|\tau_0,\Delta)& = & \frac{\Delta }{\pi}\dfrac{1}{ (\tau-\tau_0)^2+\Delta ^2}, \label{eq:g-Lorentzian}\\
		g_{Lp}(\tau|\tau_0,\Delta)&=& \frac{1}{2\Delta}\exp\left(-\frac{|\tau-\tau_0|}{\Delta}\right),\label{eq:g-Laplacian}\\ 
		g_u(\tau|\tau_0,\Delta)& = & \left\{ \begin{array}{cc} (2\Delta)^{-1},& \tau_0-\Delta\leq \tau\leq \tau_0+\Delta\\ 0,& \mbox{otherwise}\end{array} \right.,\label{eq:g-uniform}
\end{eqnarray}
representing different extremes (e.g., in terms of the asymptotics). We parameterize these DPDs via their mean/max, $\tau_0$, and variance, $\Delta$, in a similar way to facilitate comparisons. In general, we will assume that $\Delta\leq \tau_0$,
and in terms of the asymptotic analysis, we will be interested most in the regime of weak disorder, $\Delta\ll \tau_0$. (Notice that the negative values of $\tau$, $\tau<0$, are not physical.  Therefore, when performing asymptotic analysis for the disorder distributions with formally defined infinite support, described by Eqs.~(\ref{eq:g-Gaussian},\ref{eq:g-Lorentzian},\ref{eq:g-Laplacian}), one needs to make sure that the fictitious $\tau<0$ regime does not contribute the asymptotic results.) 
Then, the following averaged over the DPDs 
\begin{eqnarray}
    && \overline{P_{\uparrow,\downarrow}(x|t,\tau_0,\Delta,r)}\doteq\int d\tau g(\tau) P_{\uparrow,\downarrow}(x|t,\tau,r),
    \label{eq:disorder_average}\\
    && \overline{{\cal N}_{\uparrow,\downarrow}(t,\tau_0,\Delta,r)}\doteq\int d\tau g(\tau) {\cal N}_{\uparrow,\downarrow}(t,\tau,r),
    \label{eq:disorder_average_N}
\end{eqnarray}
will be the focus of our analysis of the inhomogeneous ensembles represented by Eqs.~(\ref{eq:g-Gaussian},\ref{eq:g-Lorentzian},\ref{eq:g-Laplacian},\ref{eq:g-uniform}).
%of the Gaussian (\ref{eq:g-Gaussian}) and Lorentzian (\ref{eq:g-Lorentzian}) types, where $g$ should be substituted by $g_G$ and $g_L$ respectively. 

In this manuscript, we pose and answer the following two related questions:
\begin{itemize}
    \item \underline{Qualitative Question about the Inhomogeneous Ensemble:} Does the disorder accelerate or slow down mixing, i.e., relaxation of the ensemble probability distribution to its steady state?
    \item \underline{Quantitative Question about the Inhomogeneous Ensemble:} How does the relaxation look depending on the system parameters and the parameters characterizing the PDF of the disorder?
\end{itemize}

Our choice of the basic model in Eq.~(\ref{eq:basic_x_model}), resulting in analytic expression for $P(x|t,\tau,r)$ stated in terms of the explicit spectral series in Section \ref{sec:basic}, allows us in Section \ref{sec:disorder} to answer the quantitative question explicitly and then to use the analytic solution to reach qualitative conclusions. 

\section{Analytic Solution for the Basic Homogeneous Model}
\label{sec:basic}

The solution of Eq.~(\ref{eq:FP_basic}) can be written in terms of the following explicit spectral expansion:
\begin{eqnarray}
        && P(x|t,\tau,r)=\sum_{k=-\infty}^{+\infty} \left(a_{k;-}(\tau,r)\xi_{k;-}(x|\tau,r) e^{-\lambda_{k;-}(\tau,r) t}+a_{k;+}(\tau,r)\xi_{k;+}(x|\tau,r) e^{-\lambda_{k;+}(\tau,r) t}\right),\label{eq:eig_exp}\\
        && \L \xi_{k;\pm}(x|\tau,r)=-\lambda_{k;\pm}(\tau,r)\xi_{k;\pm}(x|\tau,r),\label{eq:spectral}\\
        && \lambda_{k;\pm}(\tau,r)\doteq  \dfrac{r}{2}\left (1-\dfrac{W_k\left (\pm \frac{r\tau}{4}e^{\frac{r\tau}{4}}\right )}{r\tau/4} \right ),\label{eq:eignevalues}\\
        && \xi_{k;\pm}(x|\tau,r)\doteq 
            \begin{cases}
        		\begin{pmatrix}
            		\exp \left(\frac{\tau  x (r-\lambda_{k;\pm} )}{2 (\xu-\xd)}\right )\\
            		\frac{r}{r-2 \lambda_{k;\pm} } \exp\left(\frac{\tau  x (r-\lambda_{k;\pm} )}{2 (\xu-\xd)}\right)
        		\end{pmatrix},~ x<\xd\\\\
        		\begin{pmatrix}
            		\exp\left(\frac{\tau (r \xd-\lambda_{k;\pm}    x)}{2 (\xu-\xd)}\right)\\
            		\frac{(r-2 \lambda_{k;\pm} ) }{r}\exp \left(\frac{\tau  (r \xd+\lambda_{k;\pm}  (x-2 \xu))}{2 (\xu-\xd)}\right )
        		\end{pmatrix},~ \xd<x<\xu\\\\\
        		\begin{pmatrix}
            		\exp \left( \frac{\tau  (\lambda_{k;\pm}-r )x}{2 (\xu-\xd)}\right)\exp \left(\frac{\tau  (r (\xd+\xu)-2\lambda_{k;\pm}   \xu)}{2 (\xu-\xd)}\right)\\
            		\frac{(r-2 \lambda_{k;\pm} )}{r} 		\exp \left( \frac{\tau  (\lambda_{k;\pm}-r )x}{2 (\xu-\xd)}\right)\exp \left(\frac{\tau  (r (\xd+\xu)-2\lambda_{k;\pm}   \xu)}{2 (\xu-\xd)}\right)
        		\end{pmatrix},~ x>\xu
    	    \end{cases}
        ,\label{eq:eigenvectors}
\end{eqnarray}
where Eq.~(\ref{eq:eignevalues}) solves the spectral equation $r-2\lambda_{k;\pm}=\pm (r e^{\lambda_{k;\pm}\tau/2})$, and $W_k(z)$ with $z\in \mathbb{C}$ and $k \in\mathbb{Z}$ denote all the analytic in $z$ solutions of  the Lambert-W transcendental equation, $W_k(z)e^{W_k(z)}=z$. (The Lambert-W function is called \verb| ProductLog[k,z] | in Mathematica \cite{Mathematica}. See \cite{Corless1996} for details of the Lambert-W function analysis, including asymptotics.) 

To complete description of the spectral decomposition, one also needs to define adjoint eigenvalues of $\L$
\begin{eqnarray}
        && \L^\dagger \xi^\dagger_{k;\pm}(\tau,r) = - \lambda^\ast_{k;\pm}(\tau,r)\xi^\dagger_{k;\pm}(\tau,r),\label{eq:spectral_left}\\
        && \xi^\dagger_{k;\pm}(x|\tau,r)\doteq 
            \frac{\tau }{2 (\xu-\xd) ((r-2 \lambda_{k;\pm}^\ast ) \tau +4)}
            \begin{cases}
        		\begin{pmatrix}
            		 (r-2 \lambda_{k;\pm}^\ast ) \exp\left(-\frac{\tau  (r \xd+\lambda_{k;\pm}^\ast  (x-2 \xd))}{2 (\xu-\xd)}\right)\\
            		(r-2 \lambda_{k;\pm}^\ast )^2 \exp\left(-\frac{\tau  (r \xd+\lambda_{k;\pm}^\ast  (x-2 \xd))}{2 (\xu-\xd)}\right)
        		\end{pmatrix}, ~ x<\xd\\\\
        		\begin{pmatrix}
            		(r-2 \lambda_{k;\pm}^\ast ) \exp\left(-\frac{r \tau  \xd-\lambda_{k;\pm}^\ast  \tau  x}{2 (\xu-\xd)}\right)\\
            		r\exp\left(-\frac{\tau  (r \xd+\lambda_{k;\pm}^\ast  (x-2 \xu))}{2 (\xu-\xd)}\right)
        		\end{pmatrix}, ~ \xd<x<\xu\\\\
        		\begin{pmatrix}
            		 (r-2 \lambda_{k;\pm}^\ast )\exp\left(-\frac{r \tau  \xd-\lambda_{k;\pm}^\ast  \tau  x}{2 (\xu-\xd)}\right)\\
            		r \exp\left(-\frac{r \tau  \xd-\lambda_{k;\pm}^\ast  \tau  x}{2(\xu-\xd)}\right)
        		\end{pmatrix},~ x>\xu
        	\end{cases}
        ,\label{eq:eigenvectors_adj}
\end{eqnarray}
where $\L^\dagger $, the adjoint of $\L$, and the standard $L_2$ scalar product between two vectors $P$ and $G$ 
are defined according to
    \begin{eqnarray}
        && \L^\dagger\doteq 
        -u\partial_x \begin{pmatrix} 1 & 0\\ 0 & -1\end{pmatrix}-r\begin{pmatrix} \theta(x_\downarrow -x) & - \theta(x_\downarrow -x)\\ -\theta(x-x_\uparrow) & \theta(x-x_\uparrow)\end{pmatrix}, \label{eq:adjoint}\\
        && \left<G,\L P \right>\doteq\int (G^\ast)^\top \L P \,\d x
        =\int (\L^\dagger G^\ast)^\top  P \,\d x=  \left<\L^\dagger G, P \right>.
        \label{eq:scalar_product}
    \end{eqnarray}
It is straightforward to check that the eigenvectors, defined by Eqs.~(\ref{eq:eigenvectors},\ref{eq:eigenvectors_adj}), are normalized and orthogonal, i.e.,
$\left < \xi^\dagger_{k_1;\varsigma_1},\xi_{k_2;\varsigma_2}\right>=\delta_{k_1,k_2}\delta_{\varsigma_1,\varsigma_2}$.
Now closing the loop in Eq.~(\ref{eq:eig_exp}) and linking the $a$ coefficients there to the initial condition,
$P_0(x)\doteq P(0;x)$, one derives
\begin{equation}
 a_{k;\pm}=\langle \xi_{k;\pm}^\dagger,P_0 \rangle.
    \label{eq:a-initial}
\end{equation}

Substituting Eq.~(\ref{eq:eig_exp}) into Eq.~(\ref{eq:densities}), one discovers that ${\cal N}_{\uparrow,\downarrow}(t,\tau,r)$, i.e., the total density/proportion of devices that are switched on/off, is represented by the spectral series with only ``--'' modes contributing (coefficients of the ``+'' modes are exactly zero, i.e., $\int d x \xi_{\uparrow,\downarrow;+}=0$). 

We discuss consequences of the analysis on special features of the spectral problem, long time analysis (of the gap), and sensitivity of the asymptotic solution to the parameters in the following three subsections.

\begin{figure}[t]
            \centering
            \includegraphics[scale=0.65]{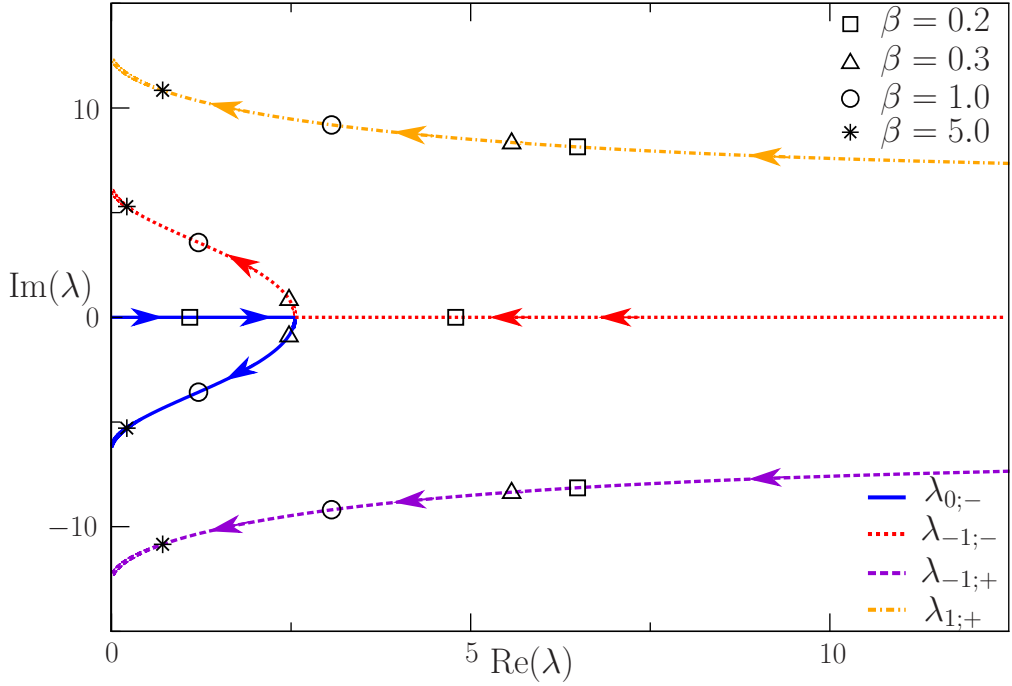}
                \caption{Evolution with increase in $\beta\doteq r\tau/4$ at the fixed $\tau=1$ of the four leading eigenvalues, $\lambda_{k;\pm}(\beta)$ (i.e., eigenvalues with the largest non-zero real value)  is shown in the $\mbox{Re}(\lambda_{k;\pm}(\beta))$-$\mbox{Im}(\lambda_{k;\pm}(\beta))$ plane. Markers indicate $\beta=0.2,0.3,1,5$. At $\beta>\beta_c\simeq 0.278$, $\lambda_{0;-}$ and $\lambda_{-1;-}$ become complex (the eigenvalues are real otherwise) and complex conjugate. The real part of $\lambda_{\pm 1;+}$ is always larger than that of $\lambda_{0;-}$ and $\lambda_{-1;-}$.}
            \label{fig:spectrum}
\end{figure}

\subsection{Features of the Spectral Problem}
\label{subsec:features}

We find it useful to identify a number of significant features of the spectral problem defined by Eqs.~(\ref{eq:eig_exp}-\ref{eq:a-initial}):
\begin{itemize}
    \item[(1)] $\lambda_{0;+}=0$ and all other eigenvalues have a positive real part that grows with $|k|$, i.e., \emph{the spectrum is discrete, positive, and ordered}.
    \item[(2)] At $r\tau> C=4W_0(e^{-1})\simeq 1.11386$, no real eigenvalues exist except the zero one, i.e., \emph{in the ``high switching rate'' regime, the solution shows oscillatory decay with time increase}.
    \item[(3)] At $r\tau\leq C$, there are only two other real eigenvalues besides zero, $\lambda_{0;-}$ and $\lambda_{-1;-}$, where $\lambda_{0;-}\leq\lambda_{-1;-}$.
    All other eigenvalues (with a nonzero imaginary part) have a real part that is larger than $\lambda_{-1;-}$.
    Therefore, \emph{in the ``low switching rate'' regime, the solution decays with time (no oscillations)}.
    \item[(4)] When one of the two parameters, $\tau$ or $r$,  is fixed, one finds that the largest value of $\lambda_{0;-}$ is achieved at the bifurcation point, where $\beta\doteq r\tau/4$ reaches $\beta_c=C/4$, i.e., \emph{given fixed $r$ and changing $\tau$}, or fixed $\tau$ and changing $r$, \emph{mixing is the fastest at $\tau=C/r$}.
    \item[(5)] Moreover, \emph{given $r$ is fixed, $d\mbox{Re}(\lambda_{0;-1})/d\tau$, i.e., the rate of change with $\tau$ of the real part of the leading eigenvalue is positive/negative when $r\tau$ is smaller/larger than $C$.}
    \item[(6)] When $r$ is fixed and $\beta=r\tau/4$ is sent to zero, one finds that $\lambda_{0;-}\to r$. Indeed, in this regime, all devices that are in the allowed range, $x\in[x_\downarrow,x_\uparrow]$, move fast to their respective boundaries; thus, \emph{at small $\tau$, PDF decay is controlled primarily by the Poisson jumps/switchings}. 
    \item[(7)] When $r$ is fixed and $\beta=r\tau/4$ is sent to $\infty$ (or alternatively when $\tau$ is fixed and $\beta$ is sent to $\infty$), one arrives at the following asymptotic:
	$\lambda_{k;\pm}\sim  (1-(\ln (\pm \beta e^\beta )+2\pi i k-\ln(\ln (\pm \beta e^\beta )+2\pi i k))/\beta)r/2$, which means, in particular,  that $\lambda_{k,-}\to 0^+-2i\pi (2k+1)/\tau$ and $\lambda_{k,+}\to 0^+-4i\pi k/\tau$. One concludes that \emph{in the asymptotic regime of the ``highest switching rate'', the temporal evolution of the PDF becomes oscillatory and relaxation to the steady state slows down asymptotically to zero.}
	\end{itemize}
Evolution of the (four) leading non-zero eigenvalues (containing the smallest real part) with the dimensionless parameter $\beta=r\tau/4$ and fixed $\tau$ is illustrated in Figure~\ref{fig:spectrum}.  
It is worth noting that the behavior of our system described by Eqs.~(\ref{eq:general_x_model}-\ref{eq:densities}) is qualitatively similar to what would be observed in a damped harmonic oscillator with the natural frequency and damping coefficient scaling respectively as $\tau$ and $1/(\tau^2 r)$.

\subsection{Long Time Asymptotic Analysis: Gap Condition}
\label{subsec:gap}
%\textcolor{blue}{ In this Section and probably in \eqref{eq:disorder_average},\label{eq:P-asymptotic-weak-disorder} and following, we should replace $\delta P, P, P^{\rm{st}}$ by $\delta \mathcal{N}_{\uparrow}, \mathcal{N}_{\uparrow}, \mathcal{N}_{\uparrow}^{\rm{st}}$. Also we have to say somewhere that at equilibrium $\mathcal{N}_{\uparrow}^{\rm{st}}=\mathcal{N}_{\downarrow}^{\rm{st}}=1/2$}

Let us now clarify the conditions under which one can limit analysis of the PDF mixing to the leading $k=0$, ``--'' mode and complex conjugate,  thus approximating
\begin{eqnarray}
&& \delta \mathcal{N}_\uparrow(t,\tau,r)\doteq  \mathcal{N}_\uparrow(t,\tau,r)-\mathcal{N}_\uparrow^{\rm{st}}(\tau,r) \approx  \exp\left(\varphi(\tau,r)-\lambda_{0;-}(\tau,r) t\right), \label{eq:P-asymptotic}\\
&& \varphi(\tau,r)\doteq \log\left(a_{0;-}(\tau,r)\int dx \xi_{\uparrow;0;-}(x|\tau,r)\right),
\label{eq:varphi_full}
\end{eqnarray}
where $\mathcal{N}_\uparrow^{\rm{st}}(\tau,r)\doteq \lim_{t\to\infty}\mathcal{N}_\uparrow(t,\tau,r)=1/2$ is the stationary solution achieved at $t\to\infty$. The approximation is valid when $\operatorname{Re}\left(\lambda_{1;+}(\tau,r)-\lambda_{0;-}(\tau,r)\right)t\gg 1$, i.e., when the relaxation time is larger than the inverse gap between real parts of the two leading eigenvalues.  
%\textcolor{blue}{The $\uparrow$ subscripts are missing from $\mathcal{N}$ and $\xi$ I guess }

$a_{0;-}$, and thus $\varphi$, depend on initial condition.  For $P_0=(\delta(x-\xd),0)$, corresponding to the ``worst case'' (least mixed) initial condition, one derives 
\begin{eqnarray}
\lambda(\tau,r)\doteq \lambda_{0;-}(\tau,r)  = && \frac{r}{2}\left (1-\frac{4 W_0\left(-\frac{r \tau e^{\frac{r \tau}{4}}}{4}\right)}{r\tau}\right ), \label{eq:lambda}
\\ \varphi(\tau,r)  = &&\log \left ( \frac{2 r (r-2 \lambda )}{\lambda  (r-\lambda ) (\tau (r-2 \lambda )+4)}\right ), \label{eq:varphi}
\end{eqnarray}
where here and below $\lambda=\lambda(\tau,r)$ is a shortcut notation for $\lambda_{0;-}$.
    
\subsection{Asymptotic Sensitivity}
\label{subsec:asymptotic}

Analytic solution, discussed above in the main body of this section, allows us to analyze the sensitivity of $\lambda_{0;-}(\tau,r)$ and $\varphi(x,\tau,r)$, defined in Eqs.~(\ref{eq:lambda},\ref{eq:varphi}), to changes in the parameter $\tau$. (The analysis can also be extended to study sensitivity to changes of $r$. We focus on the $\tau$ sensitivity because $\tau$ is user-dependent and thus uncertain, whereas $r$ is aggregator-defined and thus well controlled and certain.) Specifically, we are interested in analyzing the coefficient of Taylor expansion at $\beta_0\doteq r\tau_0/4>\beta_c$ for the dynamic characteristics of interest about (the typical) $\tau_0$:
\begin{eqnarray}
&& \varphi(\tau,r)=\varphi+(\tau/\tau_0-1)\varphi'%+\frac{(\tau/\tau_0-1)^2}{2}\varphi''
+O((\tau/\tau_0-1)^2),\label{eq:varphi_exp}\\
&& \lambda(\tau,r)=\lambda+(\tau/\tau_0-1)\lambda'%+\frac{(\tau/\tau_0-1)^2}{2}\lambda''
+O((\tau/\tau_0-1)^2), \label{eq:lambda_exp}
\end{eqnarray}
where $\varphi,\varphi',\lambda,\lambda'$ are the shortcut notations for $\varphi(x,\tau,r)$, $\tau\partial_\tau\varphi(x,\tau,r)$, $\lambda(x,\tau,r)$,
and $\tau\partial_\tau\lambda_{0;-}(\tau,r)$, respectively, evaluated at $\tau=\tau_0$. The coefficients of interest show the following asymptotics at small $\varepsilon\doteq 1/(r\tau_0)$:
\begin{eqnarray}
\lambda\tau_0 &=& -2i\pi (1-4\varepsilon+16\varepsilon^2)+16\pi^2 \varepsilon^2+ O(\varepsilon^3),\label{eq:lambda_av}\\
\lambda'\tau_0 &=& 2i\pi(1-8\varepsilon+48\varepsilon^2)-48\pi^2\varepsilon^2 +O(\varepsilon^3), \label{eq:lambda'_av}\\
\varphi &=& -\log(-i\pi)-2\pi i (\varepsilon-8\varepsilon^2)-2\pi^2\varepsilon^2+O(\varepsilon^3),\label{eq:varphi_av}\\
\varphi' &=& 2i\pi(\varepsilon-16\varepsilon^2)+4\pi^2\varepsilon^2+O(\varepsilon^3).\label{eq:varphi'_av}
\end{eqnarray}

\section{Basic Model with Disorder}
\label{sec:disorder}

Averaging over the disorder according to Eq.~(\ref{eq:disorder_average}) with only the leading $k=0$, ``--'' term in Eq.~(\ref{eq:eig_exp}) is justified when the spectral gap condition (Section \ref{subsec:gap}) is verified. Furthermore, we assume $\Delta/\tau_0\ll 1$ so that the integral Eq.~(\ref{eq:disorder_average}) is concentrated for $\tau$ located around $\tau_0$ so that $|\tau-\tau_0|\ll\tau_0$. Then, taking into account the large time asymptotic (Eq. \ref{eq:P-asymptotic}) and assuming that the Taylor series expansion (Eqs. \ref{eq:varphi_exp},\ref{eq:lambda_exp}) is legitimate (when $r\tau_0>C$), one arrives at
\begin{equation}
\left.\overline{\delta \mathcal{N}_{\uparrow}(t,\tau,r)}\right. \approx   \exp\left(\varphi-\lambda t\right)
\left.\overline{\exp\left((\tau/\tau_0-1)(\varphi'-t\lambda')
\right)}\right.. \label{eq:P-asymptotic-weak-disorder}
\end{equation}

Versions of Eq.~(\ref{eq:P-asymptotic-weak-disorder}) for the four example probability distributions of the disorder (Eqs. \ref{eq:g-Gaussian}-\ref{eq:g-uniform}) computed for small disorder $\Delta/\tau_0\ll 1$ are
 \begin{eqnarray}
\left.\overline{\delta \mathcal{N}_{\uparrow}(t,\tau,r)}\right.
\approx  \exp\left(\varphi-\lambda t\right) &\times &  \nonumber\\ 
&\mbox{(G)}:& \exp\left( \frac{\Delta^2(\varphi'-t\lambda')^2}{2}\right)\label{eq:G-weak}\\
&& \underset{\varepsilon\to 0}{\longrightarrow}
\exp\left(-2\pi^2 \frac{\Delta^2 t^2}{\tau_0^2} \right ),
\label{eq:G-weak-e0}\\
&\mbox{(Lr)}:& e^{i\Delta (\lambda' t-\varphi')}\label{eq:Lr-weak}
\\ &&\underset{\varepsilon\to 0}{\longrightarrow}\exp\left(-2\pi\frac{\Delta t}{\tau_0^2}\right),\label{eq:Lr-weak-e0}\\
&\mbox{(Lp)}:& 
\frac{1}{1-\Delta^2(\varphi'-\lambda'  t)^2}\label{eq:Lp-weak}\\ &&\underset{\varepsilon\to 0}{\longrightarrow}\frac{1}{1+(2\pi\Delta t/\tau_0^2)^2},\label{eq:Lp-weak-e0}\\
&\mbox{(u)}:& \dfrac{\sinh{\Delta (\lambda' t-\varphi')}}{\Delta (\lambda' t-\varphi')}\label{eq:u-weak}\\ && \underset{\varepsilon\to 0}{\longrightarrow}\tau_0^2\dfrac{\sin \left(2\pi\Delta  t/\tau_0^2\right)}{2\pi\Delta  t}.\label{eq:u-weak-e0}
\end{eqnarray}
The expressions are justified in their respective asymptotic limits.
% In particular in addition to the approximation made to expand at linear order around $\tau_0$ \eqref{eq:varphi_exp}\eqref{eq:lambda_exp} which gives valid result for $t\ll \tau_0^3/\Delta^2$ in the case of the Gaussian disorder Eq.~(\ref{eq:G-weak}), 
In particular, the Gaussian DPD, Eq.~(\ref{eq:G-weak}), derived via a saddle-point analysis, is valid at 
$ t \ll  \tau_0^3/\Delta^2$.
For Laplacian and uniform DPDs, we used the Laplace method %in the case of boundary maximum point 
to derive Eq.~(\ref{eq:Lp-weak}) and Eq.~(\ref{eq:u-weak}), which are valid at small disorder.
In the case of the Lorenzian DPD, we used the Cauchy integral and integration around the pole at $\tau=\tau_0-i\Delta$ of Eq.~(\ref{eq:g-Lorentzian}) to compute Eq.~(\ref{eq:Lr-weak}), again valid at small disorder. Note that in this case the slow decay of the tails in Eq.~(\ref{eq:g-Lorentzian}) makes the truncation at negative $\tau$ in Eq.~(\ref{eq:P-asymptotic-weak-disorder}) relevant only at $t\ll \tau_0^2/\Delta$. Later in time, after the $\tau_0^2/\Delta$ threshold is reached, Eq.~(\ref{eq:P-asymptotic-weak-disorder}) transitions to an asymptotic, $1/t$, decay originating from the DPD discontinuity.

\subsection{Particle Simulations and Comparison with the Theory}
\label{sec:simu}

\begin{figure}
    \includegraphics[scale=0.35]{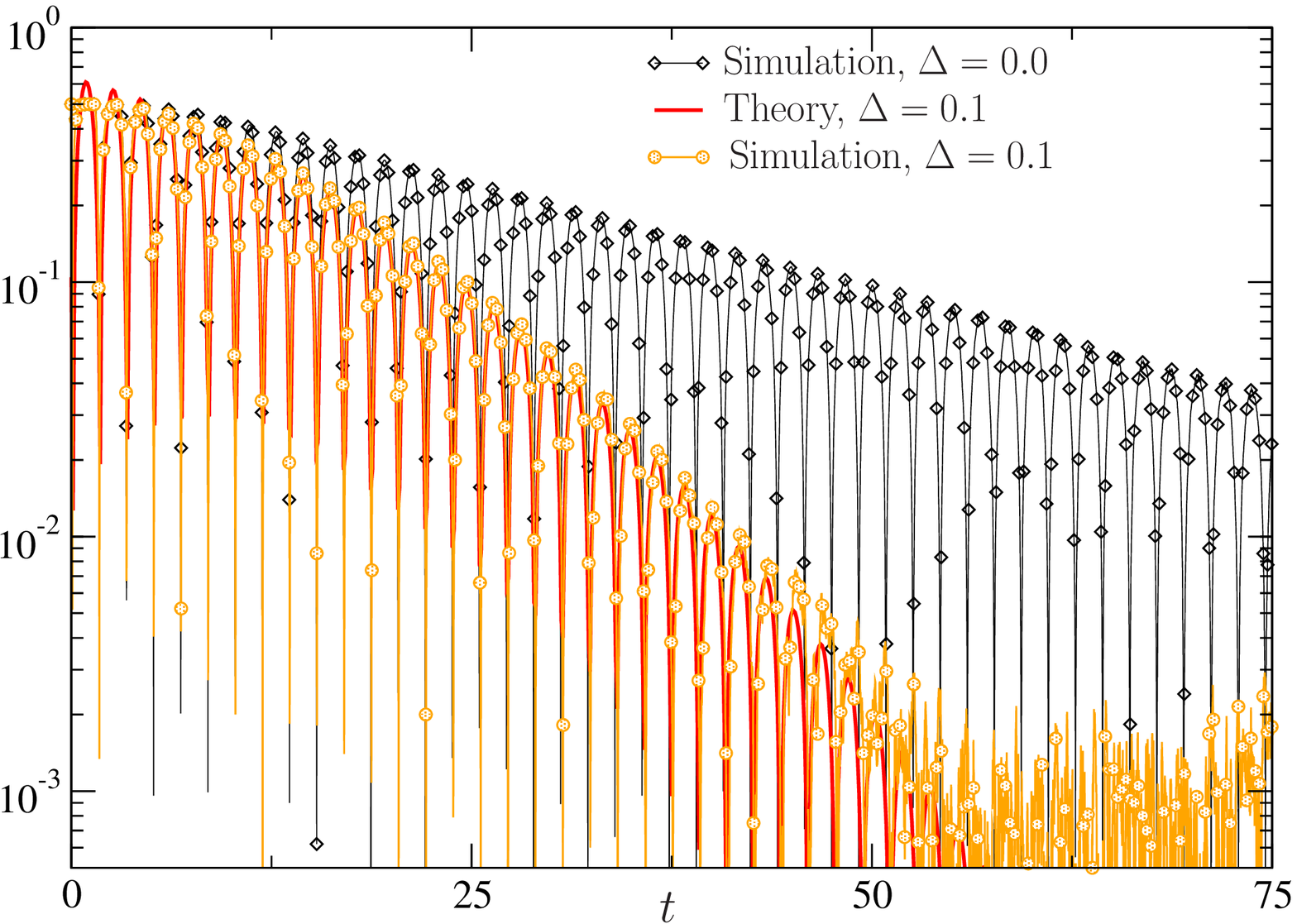}
    \includegraphics[scale=0.35]{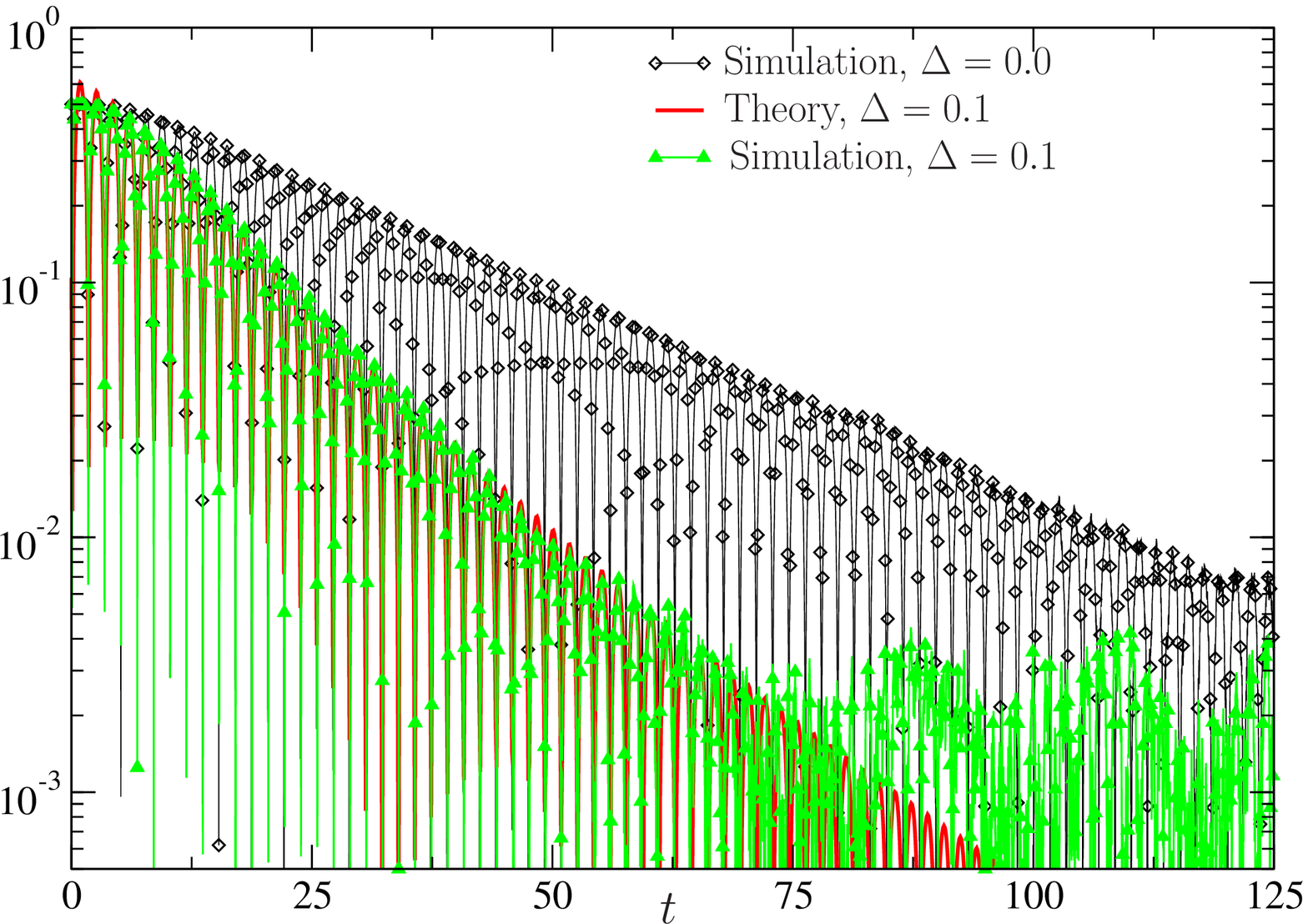}
    \includegraphics[scale=0.35]{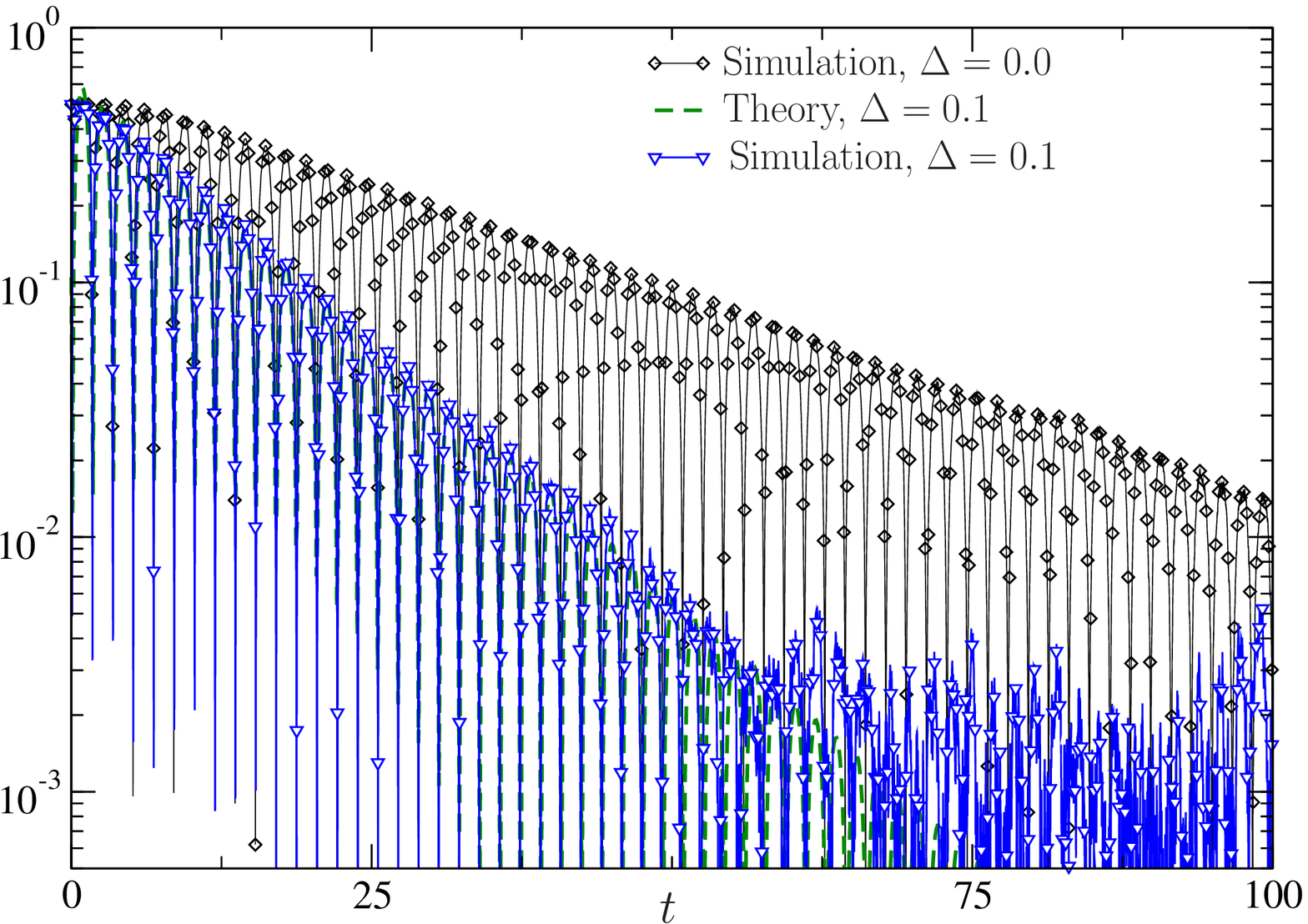}
    \includegraphics[scale=0.35]{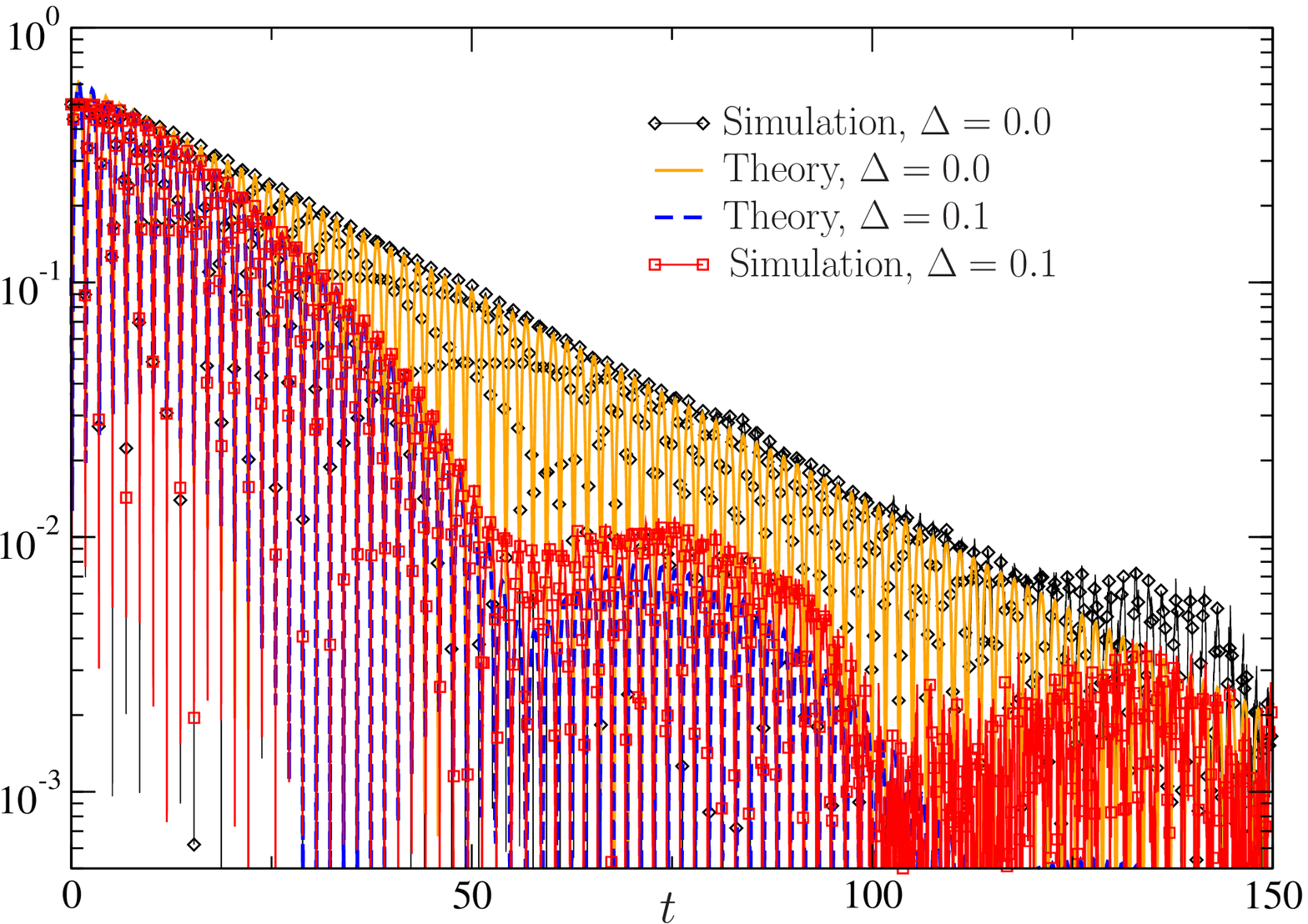}
    \caption{Time evolution of the energy consumption to its equilibrium value $|\mathcal{N}_{\uparrow}(t)-1/2|$ is shown. Four subfigures correspond to the four different DPD cases: Gaussian (top left), Laplacian (top right), Lorentzian (bottom left), and uniform finite-support (bottom right),  described by Eqs.~(\ref{eq:g-Gaussian},\ref{eq:g-Laplacian},\ref{eq:g-Lorentzian},\ref{eq:g-uniform}), respectively.
    Particle simulations are compared with analytic predictions,  with the latter limited to accounting only for
    the two dominant modes corresponding to $\lambda_{0;-},\lambda_{-1;-}$. Parameters chosen for the illustration are $r = 10$, $\tau_0 = 3$, $-\xd=\xu=1$, and $N=10^5$. Two different levels of the disorder, with $\Delta=0$ and $0.1$, are shown in each subfigure. 
    }
    \label{fig:four}
\end{figure}

To test our analytic results, we performed particle simulations of the dynamics of Eqs.~(\ref{eq:general_x_model},\ref{eq:basic_x_model},\ref{eq:sigma_model}). 
One first associates with each of $N$ devices its own relaxation time, $\tau$, drawn i.i.d from one of 
the DPDs defined by Eqs.~(\ref{eq:g-uniform},\ref{eq:g-Laplacian},\ref{eq:g-Lorentzian},\ref{eq:g-Gaussian}). (Negative values of $\tau$ are rejected.) Initially, at $t=0$, all devices are set to $x=\xd$ and $\sigma=+1$, corresponding to the ``worst case'', i.e., least mixed, initial distribution. Then the dynamics, advanced discretely and independently for each device, are implemented according to the following rules. At the beginning of each time interval, $t$, the state of each of the $N$ devices, characterized by $\sigma$ and  $x$, is advanced in time according to the first-order (in time) version of Eqs.~(\ref{eq:general_x_model},\ref{eq:sigma_model}). 
At each $t$, we monitor $\mathcal{N}_{\uparrow}(t)$, which is the total number of the devices in the state $+1$, also corresponding to the instantaneous energy consumption of the ensemble (under the model assumption that each device,  when switched on, consumes the same amount of energy.) 

The results of the straightforward particle simulations are illustrated in Figure~\ref{fig:four} for four different DPDs. To facilitate comparison, we juxtapose the results of the simulations with the corresponding analytic predictions
given by Eqs.~(\ref{eq:u-weak},\ref{eq:Lp-weak},\ref{eq:Lr-weak},\ref{eq:G-weak}). We observe  very good agreement between the theory and the simulations at short and intermediate times. The conclusion is reached based on comparison of the amplitude and the frequency of the oscillations and the relaxation rate of $\mathcal{N}_{\uparrow}(t)$.  Note  that the theory results are derived in the asymptotic, weak disorder regime described by Eq.~\eqref{eq:P-asymptotic-weak-disorder} and its complex-conjugated expression corresponding to the same (worst case) initial condition as in the simulations; hence, there are no fitting parameters. We also observe that at sufficiently large $t$, controlled by the finite (not infinite) size of the ensemble, the theory and the simulations start to deviate. Indeed, when $|\mathcal{N}_{\uparrow}(t)-1/2|$ becomes of the order of $1/\sqrt{N}$, fluctuations associated with the finiteness of the ensemble start to dominate results of the simulations. In the simulations with $N=10^5$, this threshold is reached at $|\mathcal{N}_{\uparrow}(t)-1/2|=O(10^{-3})$.

Comparing the four subfigures in Figure~\ref{fig:four} with each other is useful because it illustrates dependence of the ensemble mixing on different types of disorder. 

\section{Conclusions and Path Forward}
\label{sec:conclusion}

The main conclusion of the manuscript is that both types of randomizations, smoothing out the bang-bang control via Poisson-delayed switching 
and introducing diversity of loads in the ensemble, result in acceleration of the mixing/recovery following a heavy DR use of the ensemble. Specifically, we have shown via rigorous analysis and numerical simulations that (a) increasing the level of control (decreasing the switching rate) is advantageous only at sufficiently large rates, $r>r_c$; and (b) diversity of the devices' natural timescale (speed of cooling/heating), which is more ``regular'' (e.g., distributed according to the Gaussian DPD), is advantageous in leading to a faster mixing (more efficient recovery).

Encouraged by the reported results, we plan to extend the study in the following directions:
\begin{itemize}
    \item \underline{Complex Modeling.} We envision considering more complex models of both the individual device dynamics and the ensemble compilation.  For the former, different switching rates (for switching on and off) and more general dependence of the relaxation speed $u$ on $x$ are two practical complications  that can be included in the analysis.   For the latter (richer disorder), most significant generalization corresponds to adding disorder/inhomogeneity in other model parameters,  such as switching on/off temperatures.   
    Our working hypothesis is that these modifications/generalizations will lead to (possibly significant) quantitative but not qualitative changes in the predictions.
    
    \item \underline{Mean-Field, Nonlinear Control.} Switching rate, $r$, communicated by the aggregator to consumers, was constant in the model discussed above. It is interesting to experiment with changing the rate,  in particular allowing it to depend on the current state of the ensemble,  i.e., on the instantaneous probability distribution in the $(x,\sigma)$ space. This Mean-Field control improves greatly the relaxation time, as shown by the team in~\cite{MC2018}. 
    This intricate scenario is related to developing and extending the study to the so-called mean-field games and control \cite{huang2006}. 
    
    \item \underline{Optimal Control.} This manuscript has focused primarily on analysis of the stochastic ensemble with a control.  However, the control in this setting was not optimal but rather preset.  The natural evolution of this analysis (which would also complicate it) consists of a two-level formulation where solution of the problem analyzed here is also optimized. For example, one minimizes a cumulative cost including DR tasks (such as tracking time-evolving consumption signal from the system operator) and the mixing/recovery characteristics of the ensemble investigated above. 
    
    \item \underline{Discrete Phase Space.} Given practical constraints in the device resolution, it is natural to reduce the hybrid (continuous-discrete) state space of the analyzed model to a purely discrete space simply by binning the temperature. Moreover, following the logic of \cite{2013ETH,15PKL}, it is practically appropriate to also consider the resulting Markov Process (MP) model in discrete time. In fact, this MP formulation is also practically advantageous for analysis of the aforementioned optimal control, where the problem becomes of the Markov decision process (MDP) type, as in \cite{16BMa,16BMb,2018CCD}. We would argue that the MP and MDP approaches are naturally appropriate and algorithmically attractive to account for the randomization effects analyzed in the manuscript. 
    
    \item \underline{Data-Driven Control.} Individual devices included in the aggregation may change their behavior, which then should be accounted for through data-driven identification of a device and ensemble parameters \cite{94EM}. To track changes in real time and then account for them in the control, one would naturally resort to the data-driven approaches of the reinforcement learning type \cite{RL}, combining learning and control and aimed at developing on-line algorithms for optimal control.
\end{itemize}

\section{Acknowledgements} 

The work at LANL was carried out under the auspices of the National Nuclear Security Administration of the U.S.
Department of Energy under Contract No. DE-AC52-06NA25396. The work was partially supported by DOE/OE/GMLC and LANL/LDRD/CNLS projects. % LA-UR-18-28172

\bibliographystyle{plain}
\bibliography{bib/TCL_disorder}

\end{document}